\documentclass{phbauth}
\usepackage{graphicx}
\newcommand{\text}[1]{{\mathrm{#1}}}
\renewcommand{\vec}{\mathbf}

\begin{document}
\begin{frontmatter}
\title{Mean-field theory of magnetic properties of
Mn$_x$III$_{1-x}$V semiconductors}

\author[okl]{M. Abolfath},
\author[ind,cz,aus]{T. Jungwirth\thanksref{corresp}},
\author[ind,aus]{A.H. MacDonald}

\address[okl]{Department of Physics and Astronomy, University of Oklahoma,
              Norman, OK 73019-0225, USA}
\address[ind]{Department of Physics, Indiana University,
              Bloomington, IN 47405, USA}
\address[cz]{Institute of Physics ASCR, Cukrovarnick\'a 10, 
        162 00 Prague 6, Czech Republic}
\address[aus]{Department of Physics, The University of Texas at Austin,
              Austin, TX 78712, USA}

\thanks[corresp]{Corresponding author: fax: +1-812-855-5533;
E-mail address: jungw@gibbs.physics.indiana.edu}

\begin{abstract}
We present a mean-field theory of carrier-induced ferromagnetism
in Mn$_x$III$_{1-x}$V diluted magnetic semiconductors with a special 
emphasis placed on the magnetic anisotropy. The valence band holes are 
described using the six band Kohn-Luttinger model. We find that the
magnetic anisotropy is a complicated function of sample parameters such as
hole density or strain.  Results of our numerical simulations
are in agreement with magnetic anisotropy measurements on samples with
both compressive and tensile strains.
\end{abstract}
\end{frontmatter}

\section{Introduction}

Experiments \cite{ohno96,munekata93} in 
Mn$_x$III$_{1-x}$V  diluted
magnetic semiconductors (DMS) have demonstrated
that these ferromagnets have remarkably square hysteresis loops with
coercivities typically $\sim 40 {\rm Oe}$, and that the magnetic easy axis is
dependent on epitaxial growth lattice-matching strains.
The physical origin of the anisotropy energy in our model is spin-orbit
coupling in the valence band. Our work is based on six band envelope function
description of the valence
band holes and a mean-field treatment of their exchange interactions with ${\rm
Mn}^{++}$ ions. Even in the mean-field theory, we find that 
the magnetic anisotropy physics of
these materials is rich and we predict easy axis reorientations as a 
function of sample parameters including
hole density  or epitaxial growth lattice-matching strains. Similar
conclusions 
have been presented in a closely related, independent
study \cite{dietl}. 
A formal theory of magnetic anisotropy in cubic semiconductor DMS
is derived in Section 2. Numerical results for typical experimental samples
are presented in Section 3.

\section{Formal theory}
In zero external fields the Hamiltonian for the valence band electrons
interacting with localized $d$ electrons
on the Mn$^{++}$ ions can be written as \cite{dmsreviews}
\begin{equation}
H=H_L+J_{pd} \sum_{i,I} {\vec S_I} \cdot {\vec
s}_i \; \delta({\vec r}_i - {\vec R}_I),
\end{equation}
where $i$ labels a valence band hole and $I$ labels a magnetic ion,
$\vec S_I$ is a localized spin, $\vec s_i$ is a hole spin, and $H_L$ is
a six-band envelope-function Hamiltonian \cite{luttinger} for
the valence bands. 
In III$_{1-x}$Mn$_x$V semiconductors, the four $j=3/2$ bands are
separated by a spin-orbit splitting $\Delta_{so}$ from the two $j=1/2$ bands.
In the relevant range of hole and ${\rm Mn}^{++}$ densities, no more than four
bands are ever occupied. Nevertheless, mixing between $j=3/2$ and $j=1/2$ bands
does occur, and it can alter the balance of delicate cancellations which often
controls the net anisotropy energy. The exchange interaction between valence
band holes and localized moments is believed to be
antiferromagnetic \cite{dmsreviews}, i.e. $J_{pd}>0$. For GaAs,
experimental estimates \cite{ohno98,omiya00} of
$J_{pd}$ fall between $0.04 {\rm eV}{\rm nm}^{3}$ and $0.15 {\rm eV}{\rm
nm}^{3}$, with more recent work suggesting a value toward the lower end of this
range.

In the mean-field theory \cite{usprb,ramin} the valence band hole experience 
an effective field which depends on the average magnitude, $M$, and
orientation, $\hat{M}$, of the localized moments:
\begin{equation}
{\vec h}_{MF}(M) = J_{pd} N_{Mn} M \hat M \equiv h \hat M \label{meanfieldh}
\end{equation}
where $N_{Mn} =N_I/V$ is the number of magnetic impurities per unit volume.
The magnetic anisotropy in the absence of strain is well described by a
cubic harmonic expansion truncated at sixth order, an approximation commonly
used in the literature \cite{skomskicoey} on magnetic materials.  The
corresponding cubic harmonic expansion for total energy of a system
of non-interacting fermions
with single-particle Hamiltonian $H_{L} - h \hat M \cdot \vec s$
is
\begin{eqnarray}
E_b(\hat M) &=&E_b(\langle100\rangle) + K^{ca}_1 ({\hat
M}_x^2  {\hat M}_y^2 + {\hat M}_y^2 {\hat M}_z^2 
\nonumber \\
&+& {\hat M}_x^2 {\hat M}_z^2)
+ K^{ca}_2 \; {\hat M}_x^2 {\hat M}_y^2 {\hat M}_z^2.
\label{gammaofm}
\end{eqnarray}
The cubic anisotropy coefficients are related to total energies
for $\hat{M}$ along the high symmetry crystal directions by following
expressions:
\begin{eqnarray}
K_{1}^{ca} &=& \frac{4 (E_{b}\langle110\rangle - E_{b}\langle100\rangle)}{V}
\nonumber \\ K_{2}^{ca} &=& \frac{27 E_{b}\langle111\rangle - 36
E_{b}\langle110\rangle + 9 E_{b} \langle100\rangle}{V}. \label{cubicanisotropy}
\end{eqnarray}

The MBE
growth techniques  produce
${\rm III}_{1-x}{\rm Mn}_x{\rm V}$ films whose lattices are locked to those of
their substrates.  X-ray diffraction studies \cite{review} have established that
the resulting strains are not relaxed by dislocations or other defects, even
for thick films. Strains in the ${\rm III}_{1-x}{\rm Mn}_x{\rm V}$ film break
the cubic symmetry assumed in Eq.~(\ref{gammaofm}). However, the influence
of MBE growth lattice-matching strains on the hole bands of cubic
semiconductors is well understood \cite{valencebands} and we can use the
same formal mean-field theory to account for strain effects on
magnetic anisotropy.

\section{Numerical results}
\begin{figure}
\centerline{\includegraphics[width=8cm]{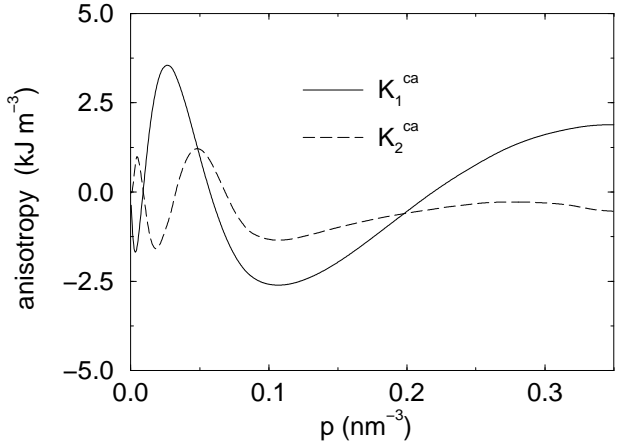}}
\caption{Cubic
magnetic anisotropy coefficients as a function of hole density}
\label{aniso}
\end{figure}

We turn now to a series of illustrative calculations intended to closely
model the ground state of ${\rm Ga}_{.95}{\rm Mn}_{.05}{\rm As}$. For this Mn
density and the smaller values of $J_{pd}$ favored by recent estimates, $h
= J_{pd} N_{Mn} J \sim 0.01 {\rm Ry}$ at zero temperature.  
This value of $h$ is not so much
smaller than the spin-orbit splitting parameter in GaAs
($\Delta_{so}=0.025$~Ry), so that accurate calculations require a six band
model.   Even with $x$ fixed, our calculations show that the magnetic
anisotropy of ${\rm Ga}_{.95}{\rm Mn}_{.05}{\rm As}$ ferromagnets is strongly
dependent on both hole density and strain. The hole density can be varied by
changing growth conditions or by adding other dopants to the material, and
strain in a ${\rm Ga}_{.95}{\rm Mn}_{.05}{\rm As}$ film can be altered by 
changing
substrates. The cubic anisotropy coefficients (in units
of energy per volume) for strain-free material are plotted as a function of
hole density in  Fig.~\ref{aniso}. 
The easy axis is nearly always determined by the leading cubic anisotropy
coefficient $K_1^{ca}$, except near values of $p$ where this coefficient
vanishes. As a consequence the easy-axis in strain free samples is almost
always either along one of the cube edge directions ($K_1^{ca} > 0$), or along
one of the cube diagonal directions ($K_1^{ca} < 0$). Transitions in which the
easy axis moves between these two directions occur twice over the range of hole
densities studied.  (Similar transitions occur as a function of $h$, and
therefore temperature, for fixed hole density.) Near the hole density
0.01~nm$^{-3}$, both anisotropy coefficients vanish and a fine-tuned isotropy
is achieved. The slopes of the anisotropy coefficient curves vary
as the number of occupied bands increases from $1$ to $4$ with
increasing hole density. This behavior is clearly seen from the correlation
between
oscillations of the anisotropy coefficients
and onsets of higher band occupations.

\begin{figure}
\centerline{\includegraphics[width=8cm]{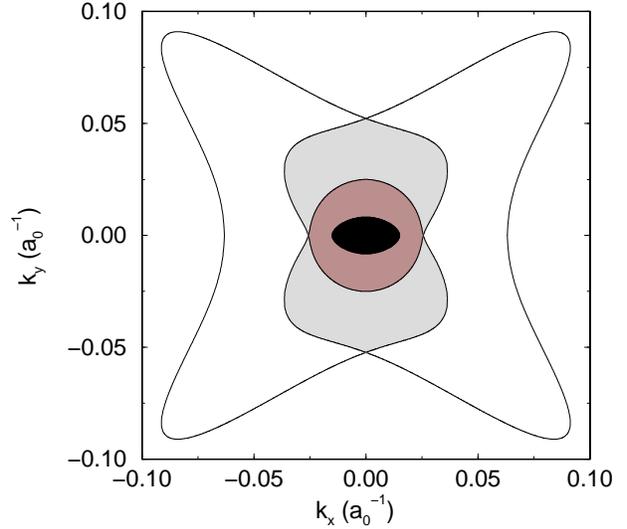}}
\caption{Six-band model
Fermi surface intersections with the $k_z=0$ plane for $p=0.1 {\rm nm}^{-3}$  
and $h=0.01 {\rm Ry}$.  This figure is for magnetization orientation is along   
the $\langle100\rangle$ direction.}
\label{nfl_6_100}
\end{figure}

\begin{figure}
\centerline{\includegraphics[width=8cm]{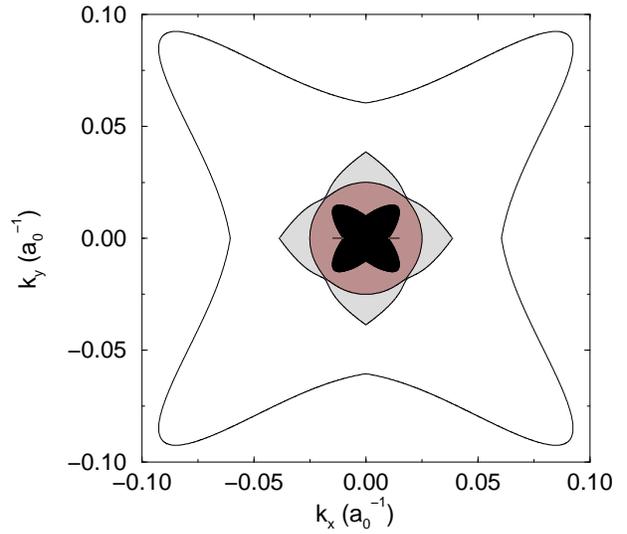}}
\caption{Six-band model
Fermi surface intersections with the $k_z=0$ plane for the parameters of
figure~\protect\ref{nfl_6_100} and magnetization orientation along the
$\langle110\rangle$ direction.}
\label{nfl_6_110}
\end{figure}

Six-band model Fermi surfaces are illustrated in
Figs.~\ref{nfl_6_100} and \ref{nfl_6_110} 
by plotting their intersections with the
$k_z=0$ plane at $p =0.1 {\rm nm}^{-3}$ for the cases of $\langle100\rangle$
and $\langle110\rangle$ ordered moment orientations.
The dependence of quasiparticle band
structure on ordered moment orientation, apparent in comparing these figures,
should lead to large anisotropic magnetoresistance effects in ${\rm
III}_{1-x}{\rm Mn}_x{\rm V}$ ferromagnets.  We also note that in the case of
cube edge orientations, the Fermi surfaces of different bands intersect.  This
property could have important implications for the decay of long-wavelength
collective modes.

\begin{figure}
\centerline{\includegraphics[width=8cm]{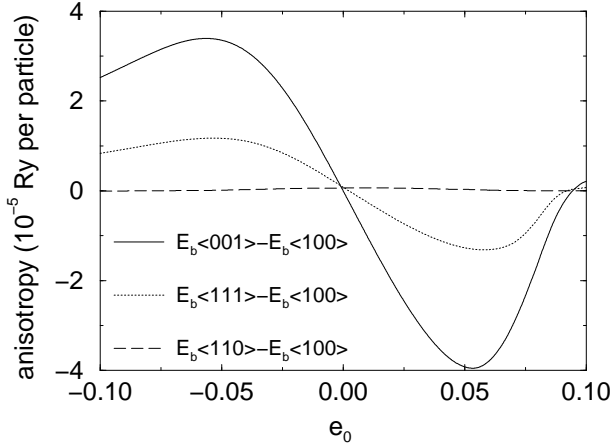}}
\caption{Energy differences among $\langle001\rangle$,
$\langle100\rangle$, $\langle110\rangle$, and $\langle111\rangle$ magnetization
orientations vs. in-plane strain $e_0$ at $h=0.01$~Ry and $p = 3.5$~nm$^{-3}$.
For compressive strains ($e_0<0$) the systems has an easy
magnetic plane perpendicular to the growth direction. For  tensile strains
($e_0>0$) the anisotropy is easy-axis with the preferred magnetization
orientation along the growth direction. The anisotropy changes sign at
large  tensile strain.}
\label{anisostrain}
\end{figure}
In Fig.~\ref{anisostrain} we present mean-field theory predictions for the
strain-dependence of the anisotropy energy at $h=0.01{\rm Ry}$ and hole
density $p=0.35 {\rm nm}^{-3}$. According to our calculations, the easy axes in
the absence of strain are along the cube edges in this case. This calculation
is thus for a hole density approximately three times smaller than the Mn
density, as indicated by recent experiments.  The relevant value of 
the in-plane strain produced by the substrate-film lattice mismatch,
\begin{equation}
e_0=\frac{a_s-a_f}{a_f},
\end{equation}
depends on the substrate on which the epitaxial ${\rm Ga}_{.95}{\rm
Mn}_{.05}{\rm As}$ film is grown.  The most
important conclusion from Fig.~\ref{anisostrain} is that strains as small
as $1\%$ are sufficient to completely alter the magnetic anisotropy energy
landscape.  For example for (Ga,Mn)As on
GaAs, $e_0=-.0028$ at $x=0.05$, the
anisotropy has a relatively strong uniaxial contribution even for this
relatively modest compressive strain, which favors in-plane moment
orientations, in agreement with experiment. A relatively small ($\sim$1 kJ
m$^{-3}$) residual plane-anisotropy remains which favors $\langle110\rangle$
over $\langle100\rangle$.  For $x=0.05$ (Ga,Mn)As on a $x=0.15$ (In,Ga)As
buffer the strain is tensile, $e_0 =0.0077$, and we predict a substantial
uniaxial contribution to the anisotropy energy which favors growth direction
orientations, again in agreement with experiment.  For the
tensile case, the anisotropy energy changes more dramatically than for
compressive strains due to the depopulation of higher subbands. 
At large tensile strains, the sign of
the anisotropy changes
emphasizing the subtlety of these effects and the latitude which exists for
strain-engineering of magnetic properties.

\section*{Acknowledgments}
We acknowledge helpful discussions with
T. Dietl, J. Furdyna,
J. K\"{o}nig, Byounghak Lee, and H. Ohno. The work was performed under
NSF grants DMR-9714055 and DGE-9902579 and Ministry of Education
of the Czech Republic grant OC P5.10.

\end{document}